\documentclass[conference]{IEEEtran}
\IEEEoverridecommandlockouts
\usepackage{cite}
\usepackage{amsmath,amssymb,amsfonts}
\usepackage{algorithm}
\usepackage{algorithmic}
\usepackage{graphicx}
\usepackage{textcomp}
\usepackage{xcolor}
\usepackage{url}
\usepackage{multirow}
\usepackage{booktabs}
\usepackage{adjustbox}
\usepackage{subcaption}
\usepackage{placeins}
\def\BibTeX{{\rm B\kern-.05em{\sc i\kern-.025em b}\kern-.08em
    T\kern-.1667em\lower.7ex\hbox{E}\kern-.125emX}}
\definecolor{greencolor}{RGB}{226,240,217}
\definecolor{bluecolor}{RGB}{222,236,248}
\definecolor{orangecolor}{RGB}{252,229,214}
\begin{document}

\title{Direct Preference Optimization for LLM-Enhanced Recommendation Systems

\thanks{$^{\star}$Corresponding author: Yunhai Tong. This work was supported by the National Key Research and Development Program of China (No. 2023YFC3807600).}
}
\author{

Chao Sun\textsuperscript{\rm 1,2,3},
Yaobo Liang\textsuperscript{\rm 1,2},
Yaming Yang\textsuperscript{\rm 1,2}, 
Shilin Xu\textsuperscript{\rm 1,2},
Tianmeng Yang\textsuperscript{\rm 1,2}, 
Yunhai Tong\textsuperscript{\rm 1,2,3}$^{\star}$\\
\IEEEauthorblockA{ \textsuperscript{\rm 1}School of Intelligence Science and Technology, Peking University}\\
\vspace{-0.5cm}
\IEEEauthorblockA{\textsuperscript{\rm 2}National Key Laboratory of General Artificial Intelligence, Peking University}\\
\vspace{-0.5cm}
\IEEEauthorblockA{\textsuperscript{\rm 3}Peking University Library}\\
\vspace{-0.5cm}
\IEEEauthorblockA{\{sunc,youngtimmy,yhtong\}@pku.edu.cn, \{yaobo.liang,yamingyang,xushilin\}@stu.pku.edu.cn}
}

\maketitle

\begin{abstract}
Large Language Models (LLMs) have exhibited remarkable performance across a wide range of domains, motivating research into their potential for recommendation systems. Early efforts have leveraged LLMs' rich knowledge and strong generalization capabilities via in-context learning, where recommendation tasks are framed as prompts. However, LLM performance in recommendation scenarios remains limited due to the mismatch between their pretraining objectives and recommendation tasks, as well as the lack of recommendation-specific data during pretraining.
To address these challenges, we propose DPO4Rec, a novel framework that integrates Direct Preference Optimization (DPO) into LLM-enhanced recommendation systems. First, we prompt the LLM to infer user preferences from historical interactions, which are then used to augment traditional ID-based sequential recommendation models. Next, we train a reward model based on knowledge-augmented recommendation architectures to assess the quality of LLM-generated reasoning. Using this, we select the highest- and lowest-ranked responses from $N$ samples to construct a dataset for LLM fine-tuning. Finally, we apply a structure alignment strategy via DPO to align the LLM’s outputs with desirable recommendation behavior.
Extensive experiments show that DPO4Rec significantly improves re-ranking performance over strong baselines, demonstrating enhanced instruction-following capabilities of LLMs in recommendation tasks.
\end{abstract}


\section{INTRODUCTION}
\label{sec:intro}
\FloatBarrier 
\begin{figure}[t]
\begin{adjustbox}{center}
    \centering
    \begin{subfigure}[t]{0.54\columnwidth}
         \centering
         \includegraphics[width=\columnwidth]{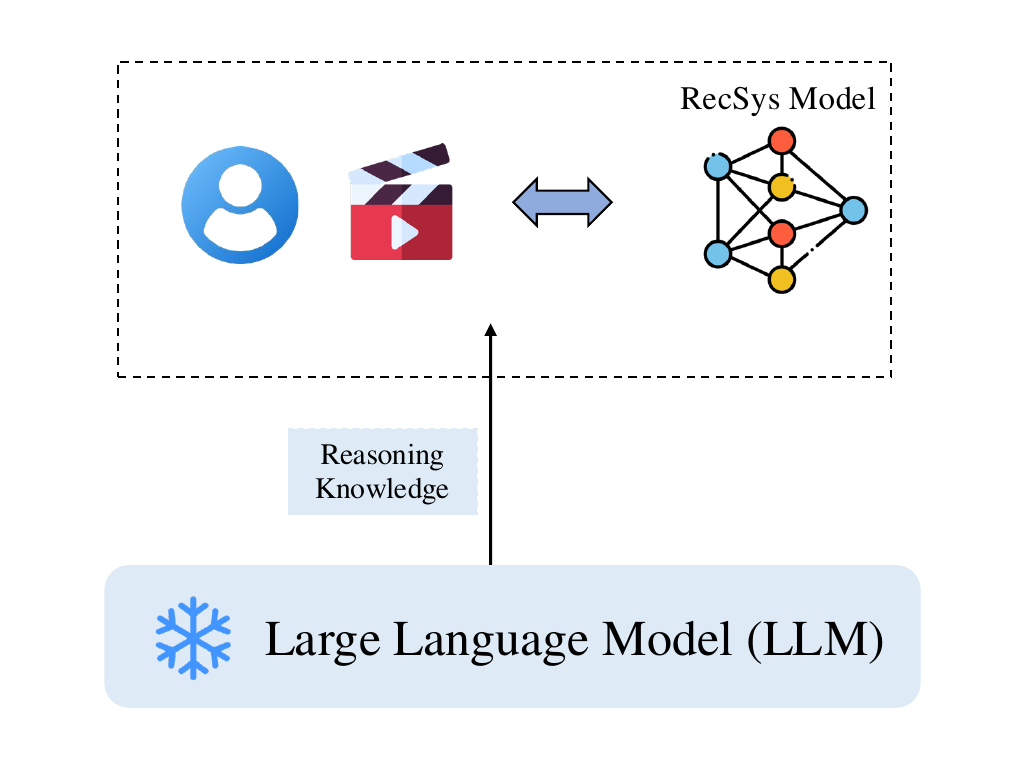}
         \vspace{-0.5cm} 
         \caption{}
         \vspace{-0.05cm}
     \end{subfigure}
     \hfill
     \begin{subfigure}[t]{0.54\columnwidth}
         \centering
         \includegraphics[width=\columnwidth]{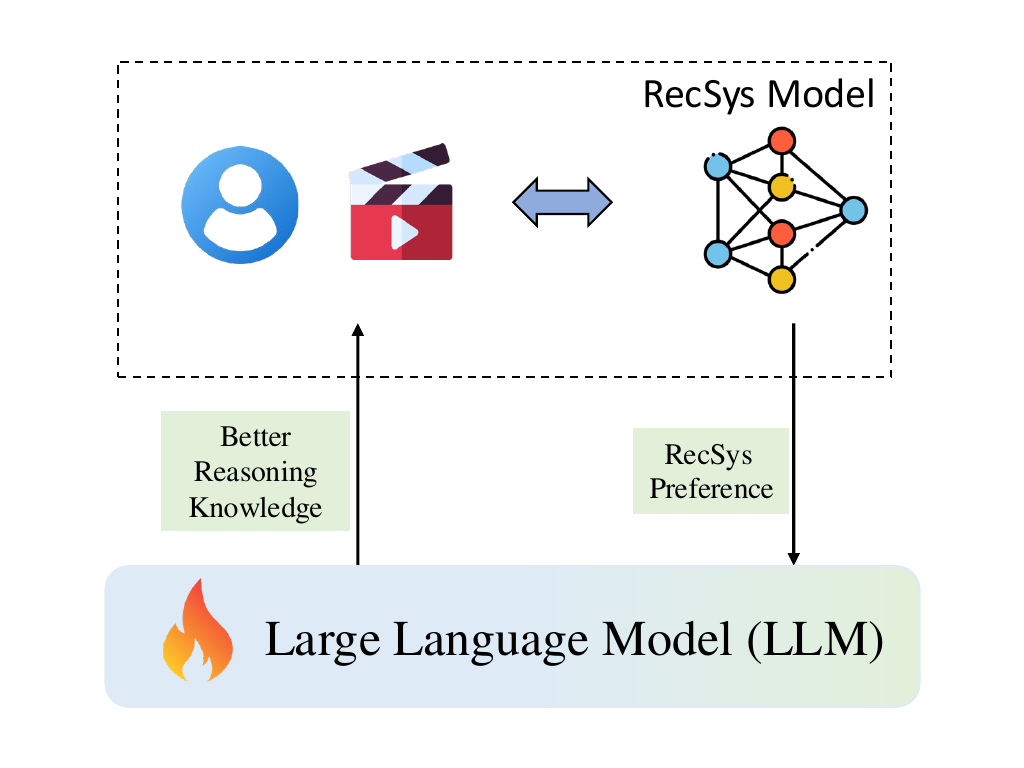}
         \vspace{-0.5cm}
         \caption{}
         \vspace{-0.05cm}
     \end{subfigure}
\end{adjustbox}
\caption{Comparison between (a) Unidirectional LLM-enhanced recommendations and (b) Bidirectional LLM-enhanced recommendations.}
\label{fig2}
\end{figure}

The rise of large language models (LLMs) is marked by their advanced natural language understanding capabilities, which enable them to comprehend and generate human-like text with high accuracy\cite{zhao2023survey,brown2020language}. These models exhibit impressive reasoning abilities, allowing them to perform complex cognitive tasks and provide insightful responses. A significant aspect of their impact is integrating traditional recommender systems with these large models, enhancing the effectiveness of recommendation processes by leveraging LLMs' extensive knowledge and contextual awareness\cite{zhao2023recommender,li2024survey}. 


However, due to the extensive world knowledge of large language models (LLMs), the augmented data they generate tends to be richer and more diverse than raw data. As illustrated in Fig.~\ref{fig2}(a), current knowledge-augmented recommendations from LLMs are implemented using zero-shot instruction tuning, which results in suboptimal performance for recommendation tasks. This suboptimality arises from the significant gap between LLM training objectives and recommendation-specific tasks, as well as the limited availability of recommendation data during pre-training. Additionally, as shown in Fig.~\ref{fig2}(b), LLM-based recommendations often overlook the feedback signals from the recommender system, which could provide valuable bidirectional benefits.


Thus, we present DPO4Rec, a framework for aligning an LLM with recommender preference for recommendations, drawing inspiration from the Reinforcement Learning from Human Feedback (RLHF) framework\cite{bai2022training}. To capture the feedback signal from the LLM, we design a reward model that evaluates the reasoning quality of recommendations enhanced by the LLM. We then select the highest- and lowest-ranked responses from the $N$ samples to construct a fine-tuning dataset, facilitating a more targeted LLM adaptation. Finally, we propose a structure alignment strategy using DPO, enabling the extraction of high-quality reasoning knowledge that significantly enhances the LLM’s relevance to recommendation tasks.
Our contributions are summarized as follows:
\begin{itemize}
\item We explore a new challenge in recommendation—aligning LLMs with knowledge-augmented recommendation models, highlighting the limitations of In-context Learning-based approaches.
\item We are among the first to leverage a recommendation model to provide feedback to large language models, aligning them with the specific goals of the recommendation task.
\item Extensive experiments across three datasets of varying sizes validate the effectiveness of our proposed DPO4Rec. Further analysis reveals that DPO4Rec generates meaningful reasoning with minimal computational cost, while preserving strong generalizability.
\end{itemize}

\section{RELATED WORK}
\subsection{LLM-enhanced Reranking in Recommendations}
Reranking has emerged as a pivotal post-processing technique in recommendations, where supplementary criteria are introduced to refine the initial ranking of candidate items. For example, DLCM\cite{ai2018learning}employs a recurrent neural network to process candidate items in a sequential manner, thereby capturing contextual dependencies. Similarly, PRM\cite{pei2019personalized} utilizes the Transformer architecture for encoding, facilitating the modeling of intricate inter-item relationships. PRM further improves its effectiveness by integrating personalized encoding strategies, such as pre-trained embeddings. SetRank\cite{pang2020setrank} employs self-attention to learn permutation-equivariant representations for input items, enhancing set modeling through invariant feature interactions. The recent advancement of Large Language Models (LLMs), like  ChatGPT and LLaMa, has triggered a new wave of interest in involving LLMs in recommender systems. Existing methods can be divided into two lines. The first is leveraging LLMs to make direct recommendations\cite{geng2022recommendation,bao2023tallrec,dai2023uncovering,geng2023vip5,li2023gpt4rec,hou2024large,liu2023chatgpt,lu2024aligning}. Typical approaches in the literature involve reformatting recommendation tasks — such as item reranking or click-through rate (CTR) prediction — into natural language constructs to facilitate the fine-tuning of LLMs\cite{lu2024aligning}. The second uses LLMs to augment traditional recommendation models with enriched user/item representations \cite{luo2024kellmrec,xi_towards_2023,liu2023first,li2023taggpt,wang2024large,ren2024representation}, which transfers the semantic knowledge from LLMs to collaborative models by aligning their latent representations, aiming to improve the recommendation performance of existing collaborative models.  
\subsection{Preference Alignment of LLMs}
The success of large language models like ChatGPT is driven not only by scaling laws but also by Reinforcement Learning from Human Feedback (RLHF)\cite{bai2022training}, which refines model outputs to align with human preferences. In addition to RLHF, Reinforcement Learning from AI Feedback (RLAIF)\cite{lee2024rlaif}has gained attention as a way to leverage AI-generated feedback for further model refinement. RLAIF uses machine-generated feedback to guide model training, often in scenarios where human feedback is scarce or expensive. The RLHF pipeline involves training a reward model followed by reinforcement learning (RL) optimization, which can be unstable and inefficient. Direct Preference Optimization (DPO) [34] addresses this by eliminating the fragile RL phase through a specific reward model parameterization, making it easier to implement while maintaining the performance of RLHF. 

\section{PRELIMINARIES}

In this section, we outline the recommendation task and define the necessary notations. The recommendation task is a binary classification problem over multi-field categorical data. The dataset is denoted as \( D = \{(x_1, y_1), \ldots, (x_n, y_n)\} \), where \( x_i \) represents the categorical features for the \( i \)-th instance and \( y_i \) denotes the corresponding binary label that indicates a click (1) or no click (0). Typically, \( x_i \) consists of user features (e.g., User ID, gender, age) and item features (e.g., item ID, gene). The feature can be represented as \( x_i = [x_{i,1}, x_{i,2}, \ldots, x_{i,f}] \), with \( f \) being the number of fields.

Recommendation models aim to learn a function \( f \) with parameters \( \theta \) to predict the click probability \( P(y_i = 1 | x_i) \), i.e., \( y_i = f(x_i; \theta) \). To meet the requirements of large language models, we follow the instruction prompt used in \cite{xi_towards_2023}, which involves extracting textual format recommendation features \( x_{i,f} \)from recommendation features \( x_i \) and organizing it into a ``Task Instruction''.

\section{METHODOLOGY}
\begin{figure*}[ht]
\centering
\includegraphics[width=1.0\linewidth]{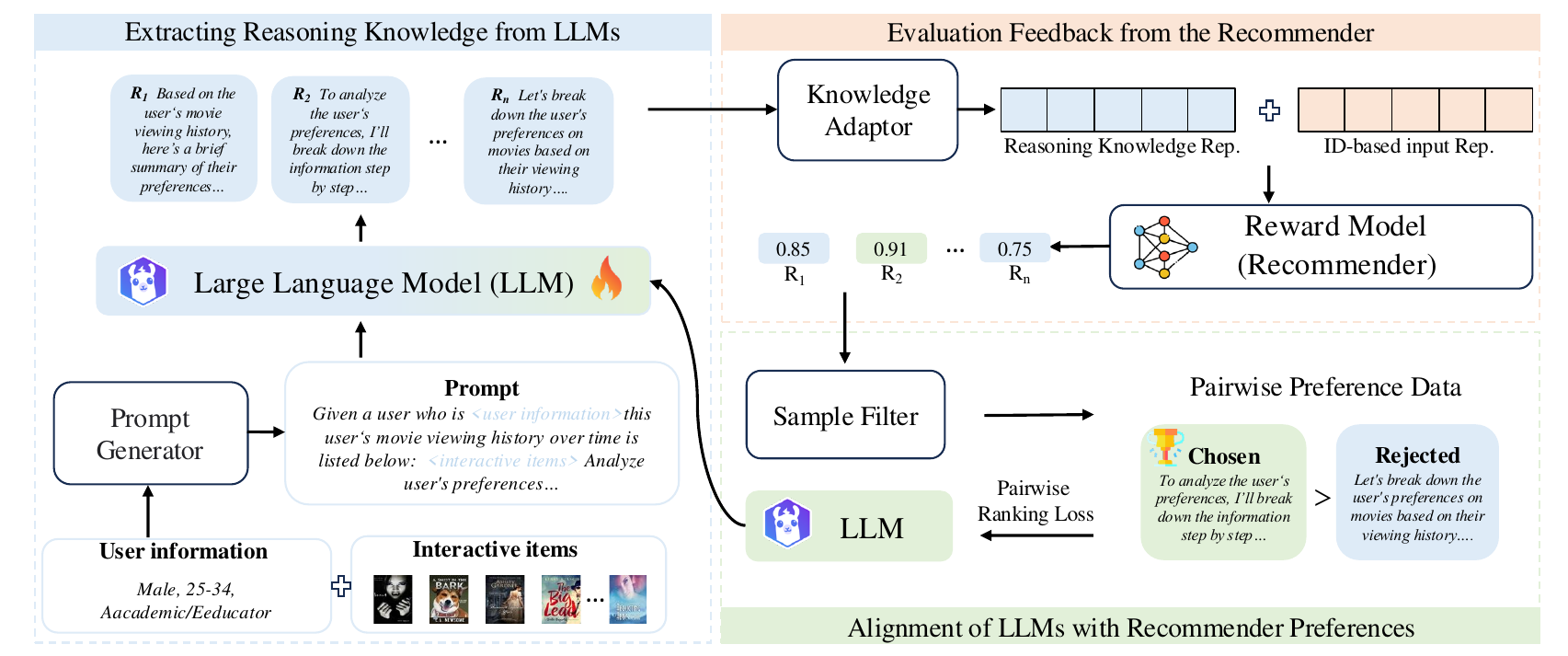}
\caption{The overview of the proposed framework.}
\label{fig3}
\end{figure*}

\begin{figure}[ht]
\centering

\includegraphics[width=\linewidth]{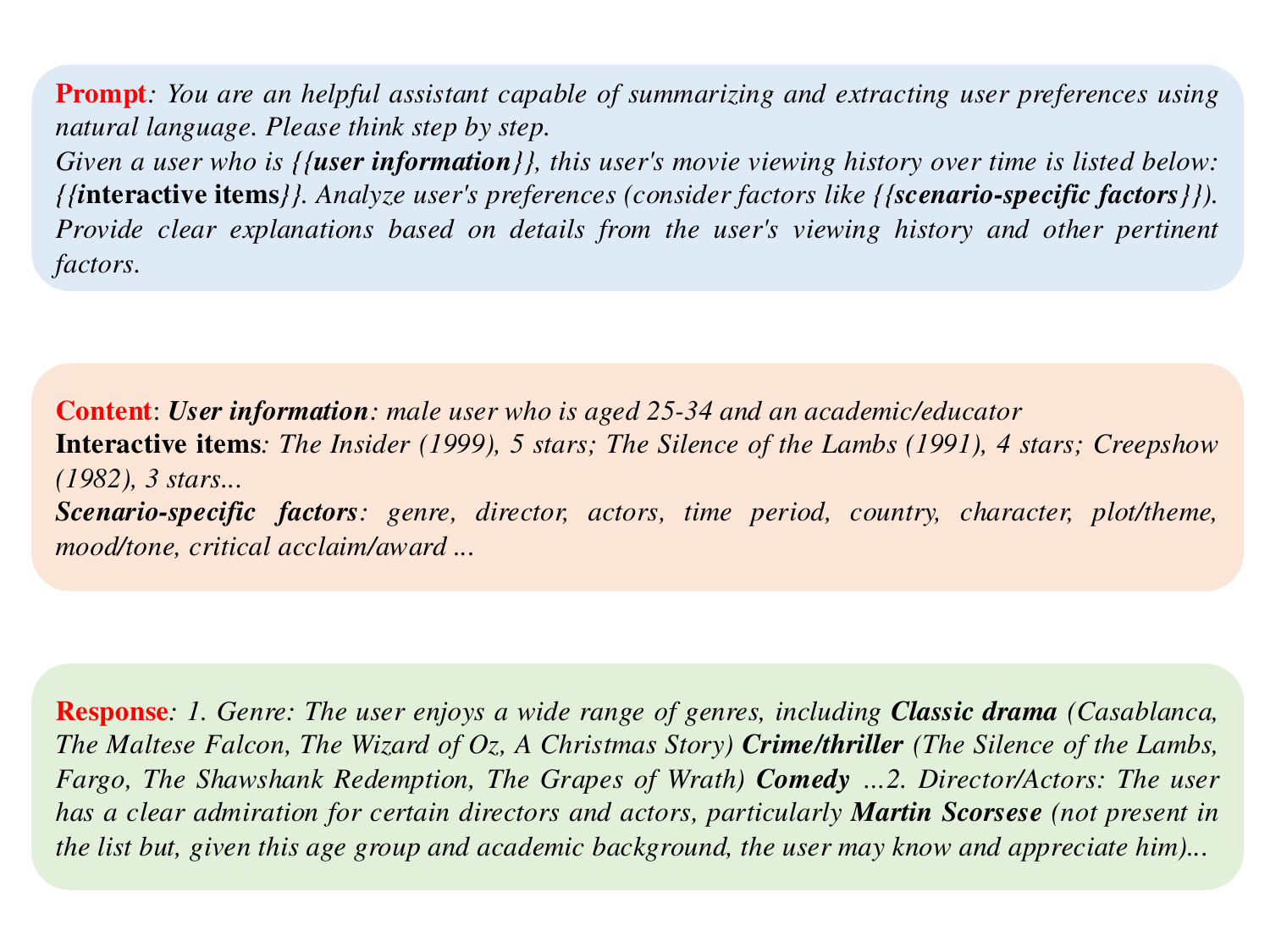}
\caption{A prompt example designed for extracting reasoning-based knowledge about user preferences from LLMs. The \colorbox{bluecolor}{blue bubble} illustrates the prompt template, which becomes a complete prompt by injecting specific user-related content (\colorbox{orangecolor}{orange bubble}) into it. The final prompt then guides the LLM in generating inferred user preferences, as represented in the \colorbox{greencolor}{green bubble}.}
\label{fig4}

\end{figure}

In this section, we will introduce our framework in detail. As shown in Fig.~\ref{fig3}, we will first introduce how to get the reasoning knowledge from the LLMs. Then, we will elaborate on how to obtain feedback on this reasoning knowledge from traditional recommendation models, and finally, we will introduce our strategy for training LLMs using this feedback.

\subsection{Extracting Reasoning Knowledge from the LLM}
Conventional recommendation methods primarily rely on explicit rating data for generating suggestions, overlooking the intermediate reasoning processes. This limitation hampers both the accuracy and explainability of the recommendations. Notably, the emergence of large language models (LLMs) has significantly advanced the ability to understand human reasoning and generate reasoning insights for recommendations. Therefore, we decompose the complex reasoning tasks by identifying significant factors determining user preferences and item characteristics. According to each factor, publicly available pre-trained models (e.g., Llama3.1) are required to generate reasoning knowledge about user preferences. Then, we meticulously designed a prompt template to facilitate reasoning according to user behavior by LLMs. Fig.~\ref {fig4} illustrates how to generate user preferences. In this case, we input an appropriate prompt text according to the required prompt template, asking the model to summarize the user's preferences with $N$ different samples. 

\subsection{Evaluation Feedback from the Recommender}
\subsubsection{Empowering Recommender with Reasoning Knowledge}
Conventional recommendation models are id-based and cannot directly utilize the reasoning knowledge of natural language forms. Thus, we utilize pre-trained language models (PLMs) to learn text representations, which allow us to measure semantic distances in vector space. Specifically, we employ the text encoder to obtain the encoding for each token within the text. Then, an adapter transforms this reasoning knowledge into low-dimensional vectors, making it compatible with traditional recommendation models. At last, we explore a straightforward approach where these augmented vectors are directly treated as additional input features.

\subsubsection{Reward Model}
 We use reinforcement learning to align the recommendation model with the capability of the LLM to generate user preferences. Unlike Reinforcement Learning from Human Feedback (RLHF), which relies on human preferences to train a reward model, we design a reward model based on the reasoning knowledge-empowered recommender to evaluate the quality of the reasoning knowledge generated by the LLM through the performance of the recommendations. Specifically, we use metrics that can be measured for individual users, such as NDCG, to calculate the score of each reasoning knowledge sample. Finally, we get a ranking list for each user consisting of the $N$ samples generated by the LLM.

\subsection{Alignment of LLMs with Recommender Preferences}
\subsubsection{Fune-tuning LLM with Pairwise Preference Data}
Based on the reward model, the highest- and lowest-ranked responses are selected as the $chosen$ and $rejected$ examples, respectively, to form a fine-tuning dataset that captures recommendation preferences. We then apply the Direct Preference Optimization(DPO)~\cite{rafailov2024direct} method to fine-tune the LLM using these pairs, resulting in an enhanced LLM-based recommendation model. The critical innovation of DPO is to avoid explicit reward modeling. Instead, it directly optimizes the language model based on recommender preferences. To achieve this, we first define a preference objective function, directly converting human preferences into the loss function used for training.

Specifically, if the recommender prefers one output $y_1$ over another potential output $y_2$, we can use the following binary cross-entropy loss function to optimize the language model $\pi$ directly:

\begin{equation*}
\begin{split}
    L_{\text{DPO}} = -\log \sigma \left(\beta \log \frac{\pi(y_1|x)}{\pi_{\text{ref}}(y_1|x)} -\beta \log \frac{\pi(y_2|x)}{\pi_{\text{ref}}(y_2|x)}\right)
\end{split}
\end{equation*}
where $\pi_{\text{ref}}$ is the base reference model. $\sigma$ is the sigmoid function and $\beta$ is a hyperparameter that defaults to 0.01.

\subsubsection{Iterative Optimization Method}
The feedback signals from the recommender system provide valuable bidirectional benefits. Leveraging this insight, we propose an Iterative Optimization Method aimed at mutually enhancing the reasoning capabilities of the LLM and the preference modeling of the recommender system. The detailed learning procedure of this iterative optimization is outlined in Algorithm~\ref{alg:algorithm1}.

\FloatBarrier
\begin{algorithm}[ht]
\caption{Iterative Optimization Method}\label{alg:algorithm1}
\begin{algorithmic}[1]
\REQUIRE Number of responses $N$, pre-trained recommendation model $Rec$, LLM with parameters $\theta_{LLM}$
\ENSURE the Final LLM-enhanced recommendation model
\STATE Generate $N$ responses using the LLM:
\FOR{$i=1$ to $N$}
    \STATE $response_i \gets LLM.generate(prompt_i)$
\ENDFOR
\STATE Evaluate each response using the pre-trained recommendation model:
\FOR{$i=1$ to $N$}
    \STATE $score_i \gets Rec.evaluate(response_i)$
\ENDFOR
\STATE Select the chosen and rejected responses:
\STATE $cho\_response \gets response_{\arg\max(score)}$
\STATE $rej\_response \gets response_{\arg\min(score)}$
\STATE Train the LLM using DPO with the chosen and rejected responses:
\STATE \parbox[t]{\dimexpr\linewidth-\algorithmicindent}{%
$LLM.train(prompt_i, cho\_response, rej\_response)$}
\RETURN Final LLM-enhanced recommendation model
\end{algorithmic}
\end{algorithm}

\section{EXPERIMENTS}
\subsection{Experiments Settings}
\subsubsection{Datasets}

Our experiments are conducted on public datasets, ML-1M\footnote{\url{https://grouplens.org/datasets/movielens/1m/}}, Amazon-Books and Amazon-Beauty\footnote{\url{https://cseweb.ucsd.edu/~jmcauley/datasets/amazon_v2/}}. ML-1M contains 1 million ratings provided by 6,040 users for 3,416 movies. Following the data processing similar to previous works\cite{zhou2018deep,xi_towards_2023}, we transform the ratings into binary labels by designating ratings 4 and 5 as positive and other ratings as negative. The data is split into training and testing sets based on user IDs, with $90\%$ assigned to the training set and $10\%$ to the testing set. The processing of Amazon-Book and Amazon-Beauty is similar to ML-1M, with the difference being the absence of user features. After filtering out users and items with few interactions, the Amazon-Book dataset contains 11,906 users, 17,332 items, and 1,406,582 interactions, while Amazon-Beauty includes 991 users, 85 items, and 5,269 interactions.

\subsubsection{Backbones and Baselines}
For the traditional recommendation models, we implement three state-of-the-art models, DLCM\cite{ai2018learning}, PRM\cite{pei2019personalized}, SetRank\cite{pang2020setrank} as backbone models. For the language model backbones, we compare GPT-4o (a large model) with three smaller models: Llama3.1-8B, Mistral-7B, and Yi-6B. At last, we chose an LLM-enhanced recommender baseline KAR\cite{xi_towards_2023}, the first practical solution that introduces logical reasoning with LLMs for user preferences to the recommendation domain.

\subsubsection{Evaluation Metrics}
Following prior works~\cite{ai2018learning,pei2019personalized,xi_towards_2023}, we adopt widely used ranking metrics including NDCG@5~\cite{jarvelin2002cumulated} and MAP@5~\cite{yue2007support} for evaluation. These metrics assess the reranking performance on the top-$K$ items selected from the complete set of items that users have not interacted with.

\subsubsection{Implementation Details}
 We utilize Llama3.1-8B\footnote{\url{https://huggingface.co/meta-llama/Meta-Llama3.1-8B-Instruct}} as the LLM of DPO4Rec and fine-tune it with LLaMA-Factory\footnote{\url{https://github.com/hiyouga/LLaMA-Factory}}, a framework designed to streamline the training of large language models. The same fine-tuning process and pipeline can be readily adapted to other LLMs. Considering a trade-off between time and payback, we chose to generate ten responses by the LLM before the re-ranking process. To obtain better performance, we follow settings from \cite{xi_towards_2023} and perform a grid search of the training configurations when training the recommendation model as the reward model. As for LLM fine-tuning, we employ the AdamW optimizer for model optimization, setting the learning rate to 5e-5, gradient accumulation steps to 8, batch size to 2, and the epoch to 3. 
\subsection{Overall Performance}
\begin{table*}[ht]
\centering
\caption{Performance comparison of models on ML-1M, Amazon-Book and Amazon-Beauty datasets.}
\label{tab:main_results}  
\begin{tabular}{l|c|cc|cc|cc}
\toprule
\multirow{2}{*}{\textbf{Backbone}} & \multirow{2}{*}{\textbf{Methods}} & \multicolumn{2}{c}{\textbf{ML-1M}} & \multicolumn{2}{c}{\textbf{Amazon-Books}}& \multicolumn{2}{c}{\textbf{Amazon-Beauty}}\\
                                 &  & \textbf{MAP@5} $\uparrow$  & \textbf{NDCG@5} $\uparrow$  & \textbf{MAP@5} $\uparrow$  & \textbf{NDCG@5} $\uparrow$  & \textbf{MAP@5} $\uparrow$  & \textbf{NDCG@5} $\uparrow$\\
\midrule
\multirow{6}{*}{\textbf{DLCM}}   & backbone           & 0.72815 & 0.77053 & 0.63823 & 0.66078  &0.79069 &0.80959\\
                                 & KAR            & 0.73482  & 0.77622  & 0.65794  & 0.68434  &0.80909 &0.82667\\
                                 & KAR$_{chatgpt-4o}$ & 0.73626  & 0.77526  & 0.66285  & 0.69002 &0.80952 &0.82935 \\
                                 & DPO4Rec$_{Yi-6B}$    & 0.73809  & 0.77822  & 0.66345  & 0.69042 &0.82496 &0.84082 \\
                                 & DPO4Rec$_{Mistral-7B}$ & 0.73666  & 0.77840  & 0.66386  & 0.69098 &0.82388 &0.84222 \\
                                 & DPO4Rec$_{Llama3.1-8B}$ & \textbf{0.73952} & \textbf{0.77942}  & \textbf{0.66563}  & \textbf{0.69310}  &\textbf{0.82734} &\textbf{0.84488}\\
\midrule
\multirow{6}{*}{\textbf{PRM}}     & backbone            & 0.73471 & 0.77558 & 0.63808 & 0.65896 &0.78759 &0.80977\\
                                 & KAR            & 0.74082  & 0.78171  & 0.66707  & 0.69481 &0.79646 &0.81408 \\
                                 & KAR$_{chatgpt-4o}$ & 0.74433  & 0.78171  & 0.66650  & 0.69440 &0.78759 &0.80977 \\
                                 & DPO4Rec$_{Yi-6B}$    & 0.74283  & 0.78329  & 0.67223  & 0.69967 &0.80620 &0.82344 \\
                                 & DPO4Rec$_{Mistral-7B}$ & 0.74280  & 0.78323  & 0.67321  & 0.70127 &0.81089 &0.82594 \\
                                 & DPO4Rec$_{Llama3.1-8B}$ & \textbf{0.74542}  & \textbf{0.78529}  & \textbf{0.67397}  & \textbf{0.70174} &\textbf{0.82035} &\textbf{0.83517} \\
\midrule
\multirow{6}{*}{\textbf{SetRank}}   & backbone           & 0.72482  & 0.76461  & 0.65314  & 0.67688 &0.74726   &0.77671 \\
                                 & KAR            & 0.73776  & 0.77828  &0.65598  &0.68317 &0.78903   &0.80977 \\
                                 & KAR\(_{chatgpt-4o}\) & 0.73626 & 0.77526 & 0.66285 & 0.69002 &0.79170   &0.81175\\
                                 & DPO4Rec\(_{Yi-6B}\)    & 0.73972 & 0.78087  & 0.65685  & 0.68291 & 0.80108   &0.82187 \\
                                 & DPO4Rec\(_{Mistral-7B}\) & 0.74167 & 0.78265 & 0.65935 & 0.68495 &0.79177   &0.81186\\
                                 & DPO4Rec\(_{Llama3.1-8B}\) & \textbf{0.74352} & \textbf{0.78430}& \textbf{0.66061} & \textbf{0.68686} &\textbf{0.81385}   &\textbf{0.83151} \\
\bottomrule
\end{tabular}

\end{table*}
\subsubsection{Improvement over Backbone Models}
We first evaluate the effectiveness of our method across three representative backbone architectures: DLCM, PRM, and SetRank. As illustrated in Table~\ref{tab:main_results}, the proposed DPO4Rec consistently demonstrates superior performance compared to the original backbone models on all evaluated datasets. Specifically, for the ML-1M dataset, the DPO4Rec$_{Llama3.1-8B}$ variant achieves the best performance across all backbones, with relative MAP@5 improvements of 1.45\%, 1.35\%, and 1.42\% over DLCM, PRM, and SetRank, respectively. Similar improvements are also evident in the Amazon-Books and Amazon-Beauty datasets, confirming the robustness and generalizability of our method. These results substantiate the efficacy of incorporating DPO4Rec optimization into existing recommendation frameworks.
\subsubsection{Improvement over Baselines}
We further compare DPO4Rec against the KAR baseline, an existing optimization approach enhanced by ChatGPT-4o. Experimental results clearly indicate that DPO4Rec variants consistently outperform KAR across all backbone models and datasets. Notably, on Amazon-Beauty with PRM as backbone, DPO4Rec$_{Llama3.1-8B}$ surpasses KAR by 3.92\% in terms of NDCG@5, reflecting significant gains attributed to the sophisticated instruction-following capability and reasoning provided by the optimized DPO4Rec model. The consistent advantage of DPO4Rec over KAR highlights the practical benefit of the proposed optimization strategy for sequential recommendation tasks.
\subsubsection{Robustness in Combining with Different LLMs}
The DPO4Rec framework demonstrates remarkable flexibility when integrated with various large language models (LLMs). As shown in Table~\ref{tab:main_results}, our proposed DPO4Rec framework consistently outperforms GPT-4o, even when paired with lightweight LLMs such as Llama3.1-8B, Mistral-7B, and Yi-6B. This highlights that the direct application of LLMs as recommenders has yet to yield satisfactory outcomes. DPO4Rec, however, effectively leverages these models to achieve superior performance, showcasing its robustness and adaptability across different LLM architectures.

\subsection{Ablation Study}
\begin{figure}
	\centering
	\begin{subfigure}[t]{0.47\columnwidth}
         \centering
         \includegraphics[width=\columnwidth]{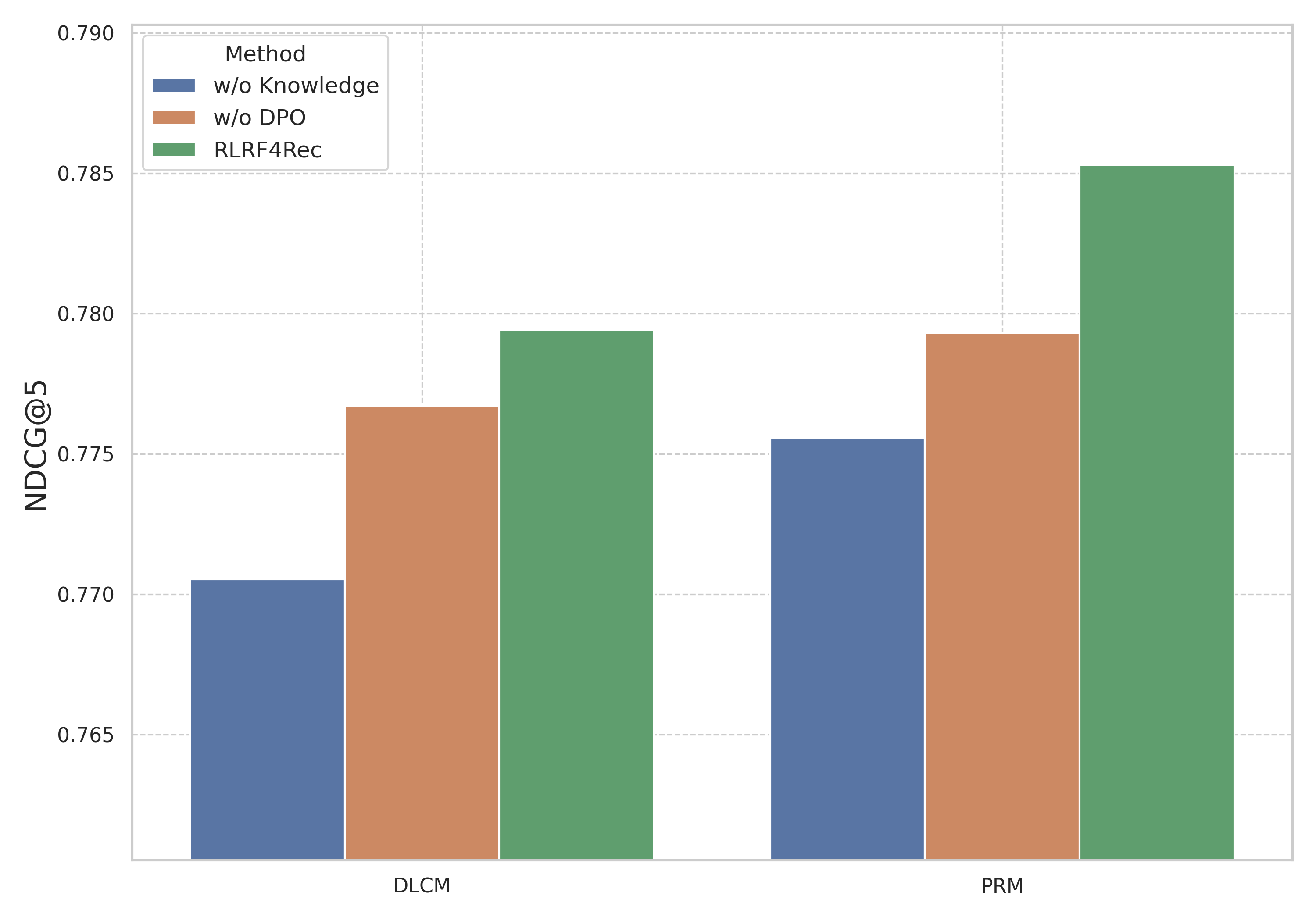}
         \vspace{-0.5cm} 
         \caption{ML-1M}
         \vspace{-0.05cm}
     \end{subfigure}
     \hfill
     \begin{subfigure}[t]{0.47\columnwidth}
         \centering
         \includegraphics[width=\columnwidth]{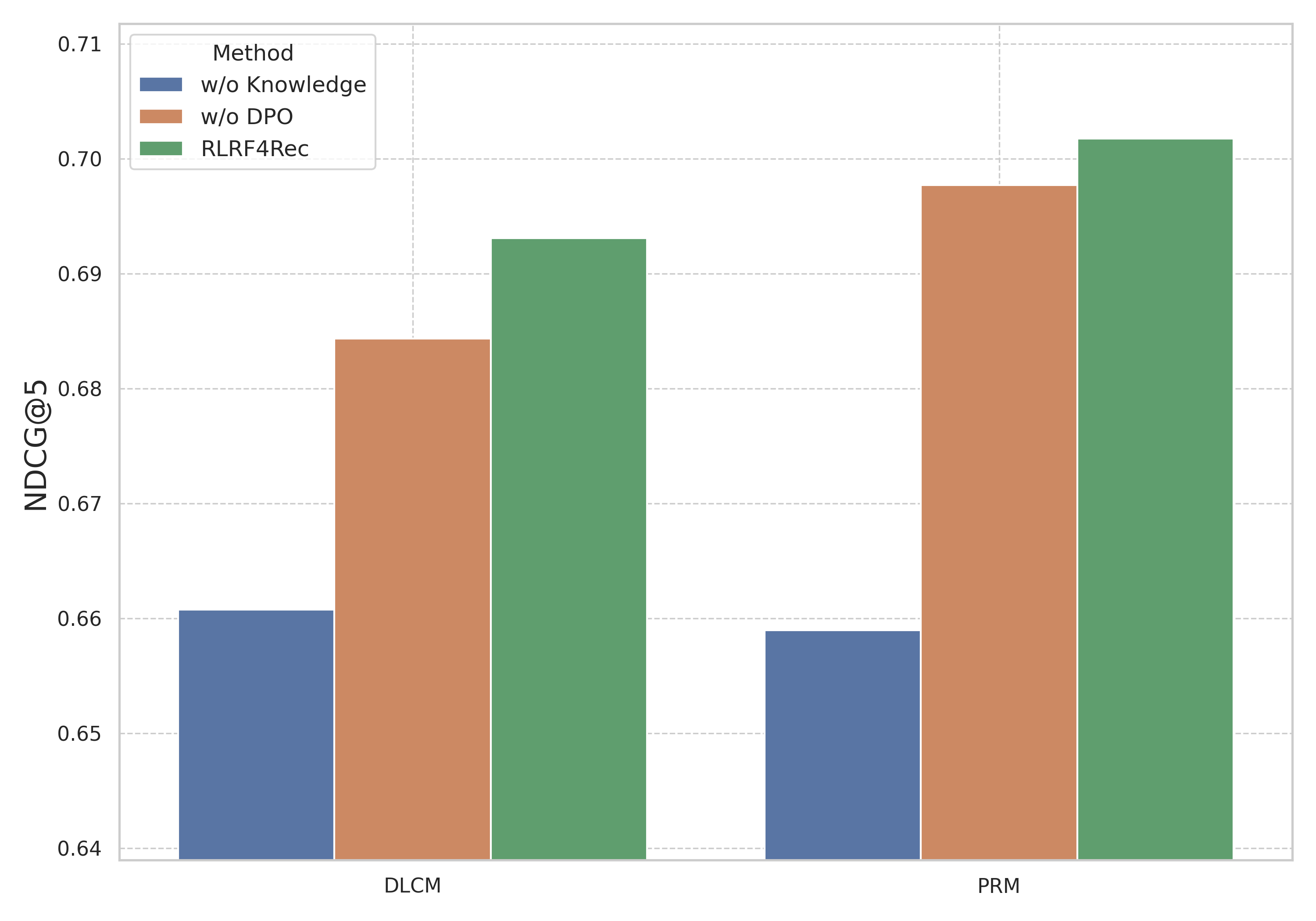}
         \vspace{-0.5cm}
         \caption{Amazon-Books}
         \vspace{-0.05cm}
     \end{subfigure}
     \vspace{-0.05cm}
       \caption{Ablation study about Knowledge and DPO}
       \label{fig:Ablation Study}
       \vspace{-0.4cm}
\end{figure}

In this subsection, we conduct experiments to analyze how different components of DPO4Rec impact its performance. As illustrated in Fig.~\ref{fig:Ablation Study}, removing the reasoning knowledge component results in a significant performance drop, particularly on the Amazon-Books dataset, highlighting its critical role in providing richer contextual insights. In contrast, omitting the DPO component leads to a relatively moderate decline, demonstrating that the combination and synergy of all components collectively drive optimal recommendation performance. Thus, reasoning knowledge primarily enhances recommendation accuracy through contextual enrichment, while the DPO fine-tuning further refines the decision-making process, boosting overall effectiveness.
\subsection{Performance with Iteration Number}
\begin{figure}
	\centering
	\begin{subfigure}[t]{0.47\columnwidth}
         \centering
         \includegraphics[width=\columnwidth]{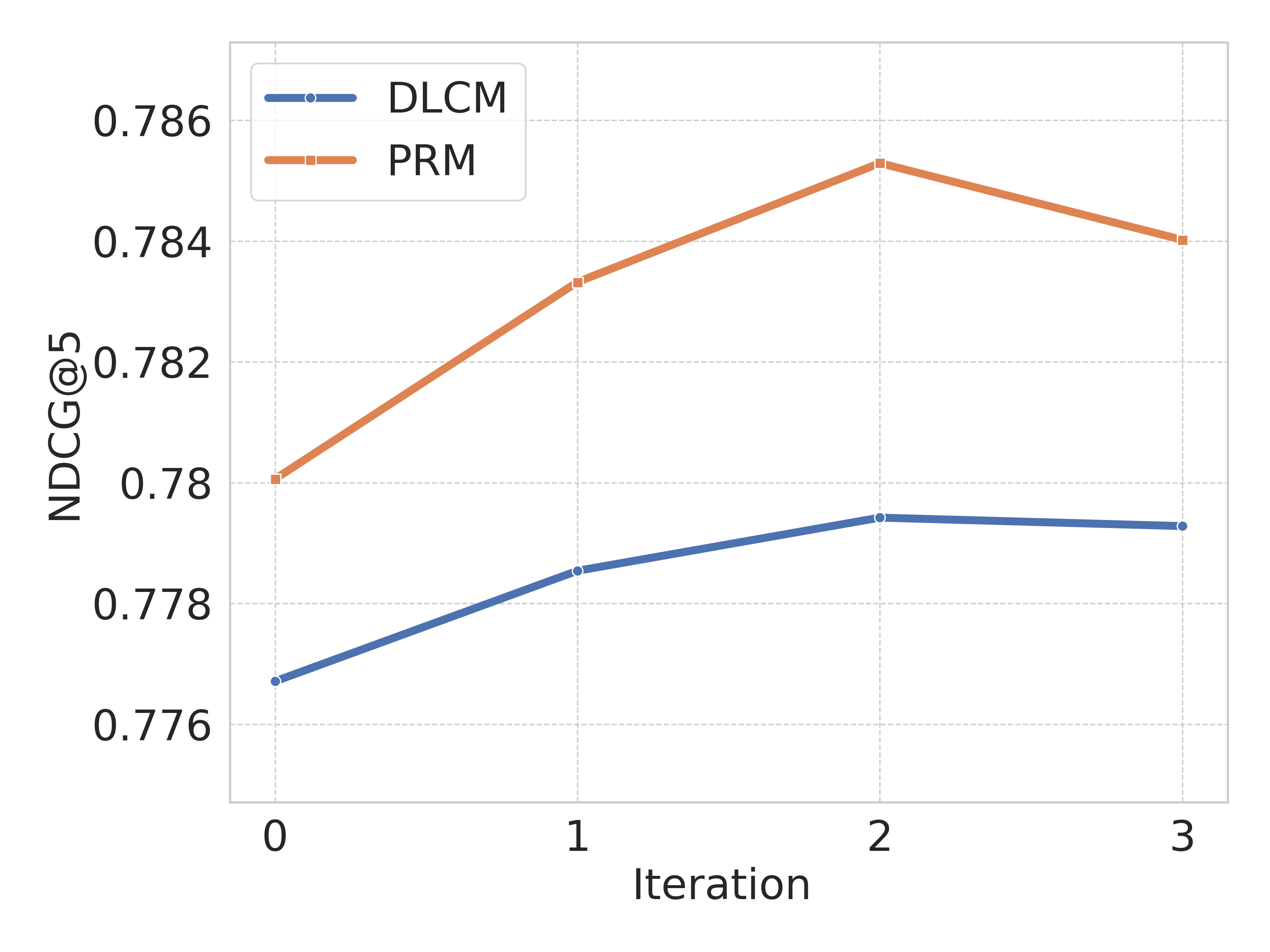}
         \vspace{-0.5cm} 
         \caption{Performance with iterations on ML-1M}

     \end{subfigure}
     \hfill
     \begin{subfigure}[t]{0.47\columnwidth}
         \centering
         \includegraphics[width=\columnwidth]{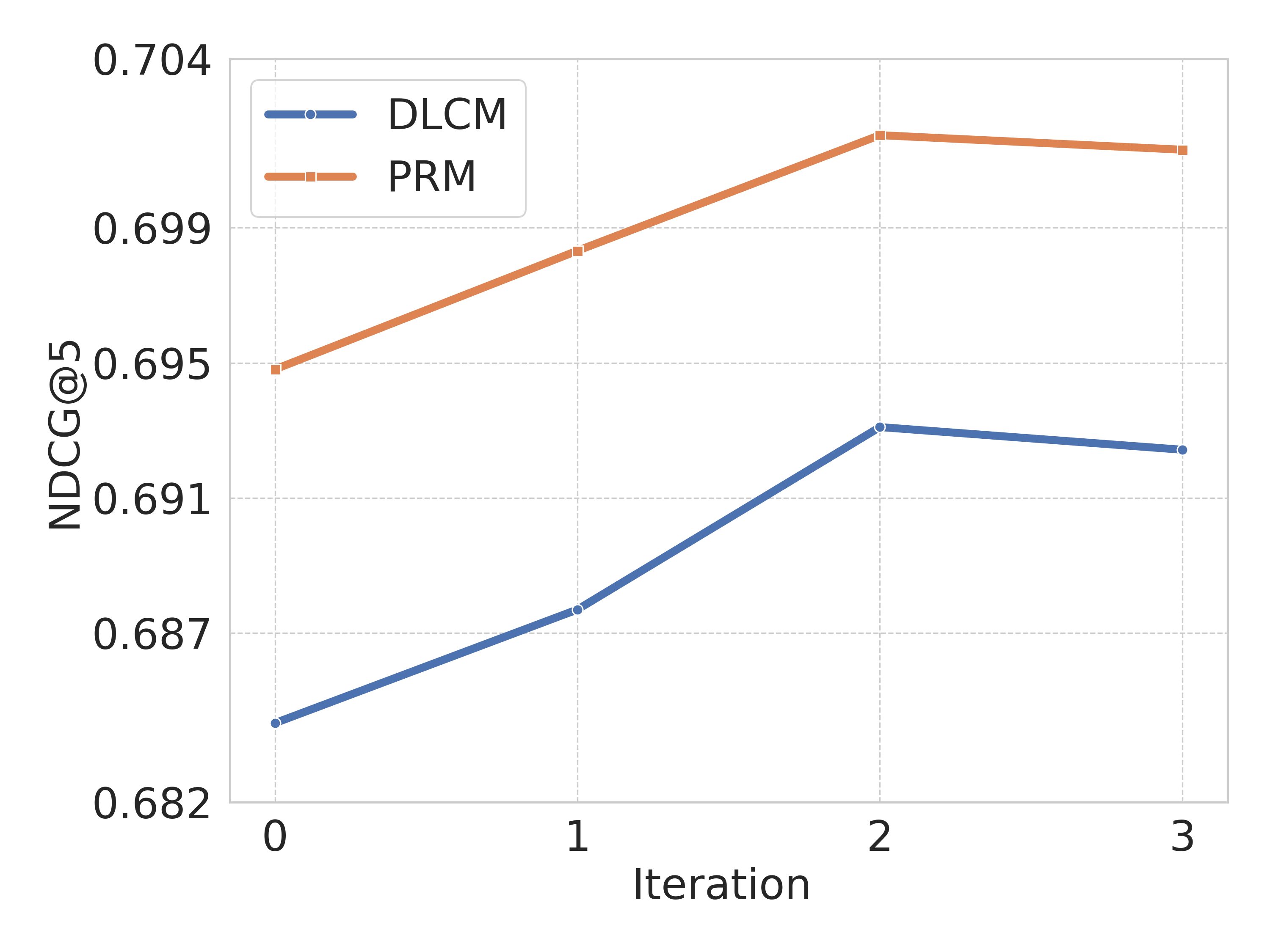}
         \vspace{-0.5cm}
         \caption{Performance with iterations on Amazon-Books}

     \end{subfigure}
     \begin{subfigure}[t]{0.47\columnwidth}
         \centering
         \includegraphics[width=\columnwidth]{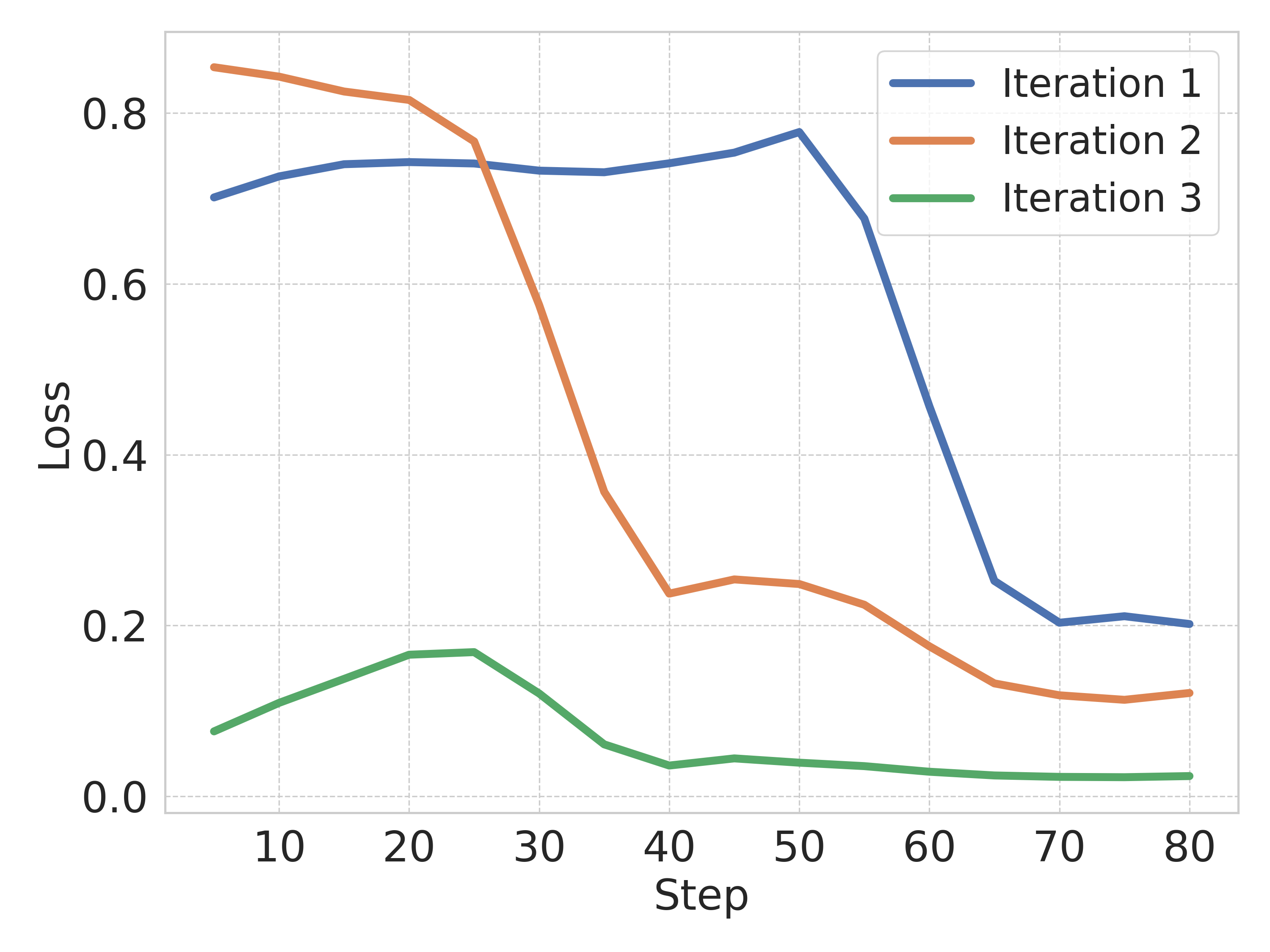}
         \vspace{-0.5cm}
         \caption{Study of Loss with iterations}

     \end{subfigure}
     \hfill
     \begin{subfigure}[t]{0.47\columnwidth}
         \centering
         \includegraphics[width=\columnwidth]{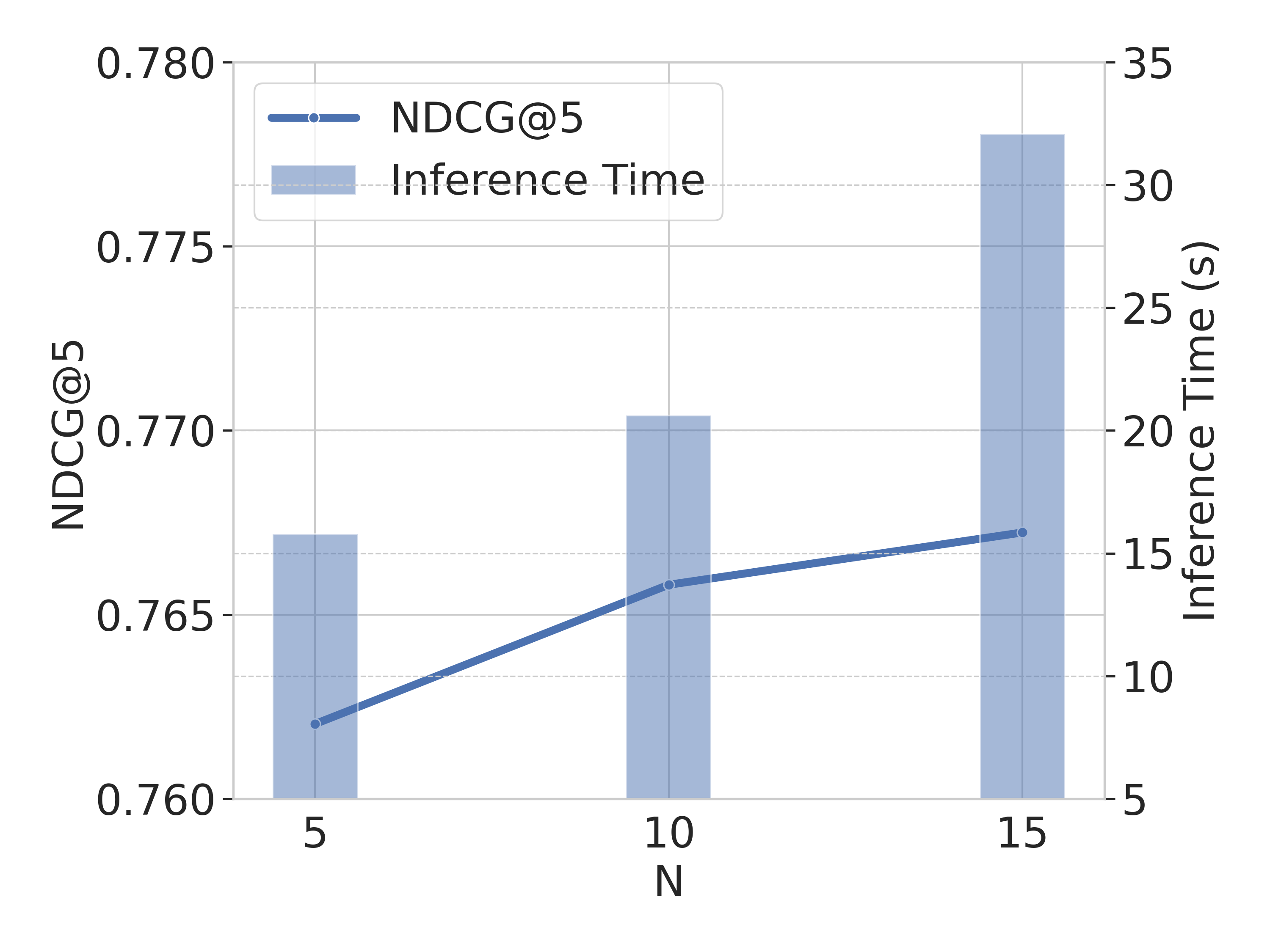}
        \vspace{-0.5cm} 
         \caption{Impact of Number of $N$}

     \end{subfigure}
     
     \vspace{-0.05cm}
       \caption{Analysis of iterations and samples.}
       \label{fig:analysis}
       \vspace{-0.4cm}
\end{figure}
We want to verify the impact of varying numbers of iterations. The results are depicted in Fig.~\ref{fig:analysis}(a)(b). We find that as the number increases, the model performance initially improves and then declines, with the optimal performance occurring when the number is set to 2. Additionally, we note that performance dropped in round 3. As shown in Fig.~\ref{fig:analysis}(c), this limitation may arise from the overfitting problem in DPO training. As the number of training epochs increases, the model may become overly aligned with the training data, which ultimately harms its performance on test data. The issue has also been discussed in previous work \cite{azar2024general}, which asserts that weak regularization and overfitting in DPO can cause a decline in policy performance. Our study strongly supports this conclusion.
\subsection{Impact of Number of Samples ($N$)}
In this subsection, we investigate the influence of the number of generated samples ($N$) on inference time and model performance. Specifically, we perform an ablation study by varying $N$ from 5 to 15, as shown in Fig.~\ref{fig:analysis}(d). The results indicate that increasing the number of samples generally leads to improved recommendation performance, but at the cost of higher inference latency. Considering this trade-off between computational efficiency and recommendation quality, we select $N=10$ samples, which offers a reasonable balance by delivering substantial performance improvements without incurring excessive inference overhead.

\section*{CONCLUSION}
In conclusion, this work proposes DPO4Rec, a novel framework that enhances the effectiveness of large language models (LLMs) in recommender systems by mitigating the gap between LLM pre-training and recommendation tasks. DPO4Rec leverages user preferences inferred by LLMs to enrich traditional sequence-based recommendation models, integrating insights from recommender feedback. The framework introduces a reward model to evaluate reasoning capabilities and employs Direct Preference Optimization (DPO) to fine-tune LLMs for improved alignment with recommendation objectives. Extensive experiments confirm the effectiveness of DPO4Rec, demonstrating substantial improvements in recommendation re-ranking metrics. The results highlight that, with appropriate guidance and alignment, LLMs can significantly enhance performance in recommendation scenarios.

\bibliographystyle{IEEEbib}
\bibliography{icme2025references}
\end{document}